\documentclass[12pt]{amsart}
\usepackage[utf8]{inputenc}
\usepackage{xcolor}
\usepackage{geometry}                		% See geometry.pdf to learn the layout options. There are lots.
\theoremstyle{definition}

\theoremstyle{plain}

\begin{document}

\title[On the topics of my 
conversations with Igor Frenkel]{On the topics of my\\ 
conversations with Igor Frenkel }

\author{Samson L.  Shatashvili}
\address{}
\address{The Hamilton Mathematics Institute, Trinity College Dublin, Ireland}
\address{The School of Mathematics, Trinity College Dublin, Ireland}
\address{Simons Center for Geometry and Physics, Stony Brook University, USA}
\address{}

\email{samson@math.tcd.ie, sshatashvili@scgp.stonybrook.edu}

%\date{May 2023}

\begin{abstract}
This is a summary of my lecture at Igor Frenkel's 70th birthday conference. I give a brief review of my almost forty years of scientific interactions with Igor.  I focus here on three topics of joint interests: 2-cocycles, coadjoint orbits and $WZW_4$. At the end of each topic I comment on new developments, if any.
\end{abstract}

\dedicatory{Dedicated to Igor Frenkel on his 70th birthday}

\maketitle

\vspace{0.5cm}
\section{Introduction}
\vspace{1cm}

Talks on such occasions are usually of two types: Type I -  speakers say a few words about the birthday person and continue to describe their own recent work (or similar); Type II - speakers give a talk specifically designed for the birthday person covering topics of their past/present joint interests. I adopted the latter, Type II, approach in my talk \cite{T}.

First time Igor and I met was around Thanksgiving in 1989 during my brief visit to Yale. But I knew Igor's work much before. In 1983 L. Faddeev and I were working on quantum anomalies and group cohomology and somewhere, in one of the western countries, Faddeev met I. M. Singer who told him that together with Igor they discovered 2-cocyle in three-dimensions but did not publish because they received (I believe from R. Jackiw) our manuscript \cite{FS} with the same result. In fact, there is a comment in I. M. Singer's Ast\'erisque article about this \cite{FrSin}. Later I learned from Igor how one early morning in spring/summer of 1984 Singer knocked on his door in Aspen to inform him about this "bad" news. The first part of this talk, Section 2, will be about this 2-cocycle.

Igor and I became colleagues in 1994 when I moved to Yale. He had been at Yale for over a decade and in addition to being a colleague he became a close friend who would guide me in the new environment. We had a joint seminar series between Physics and Mathematics Departments (interestingly, I. M. Singer was responsible for such a program to exist) - ``Geometry, Physics, Symmetry". This was one of the most successful seminar series I have ever witnessed in my forty years in academia. It very much shaped our life at Yale.  At that time I was very much interested in the work of H. Nakajima on relation between instantons on ALE spaces and Kac-Moody Algebras. I learned about Nakajima's work from C. Vafa when we were working on string theories on exceptional holonomy manifolds in the Spring 1994 (Cumrun was working  with E. Witten on $N=4$ SYM and S-Duality at the same time and he pointed me to \cite{Nak} which played an important role in their work). Since instanton moduli space was one of the topics of my PhD thesis 10 years before I obviously got very interested and shared my thoughts with Igor immediately after I arrived at Yale in the Fall of 1994. Not surprisingly it turned out that Igor was thinking about exactly the same thing, completely independently and on his own. So, we invited H. Nakajima to speak at ``Geometry, Physics, Symmetry" and at the same time asked I. Grojnowski (who had just arrived at Yale as a postdoc in department of mathematics) to deliver series of lectures on the topics very much related to Nakajima's work. All this led to a very successful set of papers for both Igor and me \cite{FrKh}-\cite{Issues}.  
%\cite{FrTod}, \cite{LMNSavatar}, \cite{LMNScoc}, \cite{LMNStwo}, \cite{MNSone}, \cite{MNStwo}, \cite{LNSone}
I devote the third part of this talk to some aspects of this large topic - there have been recent developments related to one of the spinoffs of it (mainly related to \cite{FrKh}, \cite{LMNSavatar}, \cite{LMNStwo}), due to K. Costello \cite{Costello}, and I will make a few comments about it below in Section 4.

Lastly, I want to mention coadjoint orbits, one of the topics of Igor's early interests, originating in late 70s and included in his PhD thesis \cite{Fr1}. I myself became interested in this topic much after Igor, in 1988-89, when together with A. Alekseev and L. Faddeev we were looking at path integrals (quantum mechanics) over coadjoint orbits as proper ``upgrade"  of A. Kirillov's original approach (including Kac-Moody and Virasoro coadjoint orbits \cite{AS1}, \cite{AS11}). I recall many discussions with Igor on this topic at Yale in the 90s. Of course, Igor had numerous ideas how to improve these methods and I hope in this talk I can present one recent result (which might, or might not, surprise Igor - as we indeed discussed something very related 30 years ago) obtained in joint work with A. Alekseev \cite{AS2}, \cite{ASDH}. This I discuss in Section 3. 
%here - I will explain exactly how Igor's formula from \cite{Fr1} can been obtained as a special limit of character formula. 

{\bf Acknowledgements:} 
I thank A. Alekseev for the long-standing collaboration 
%\cite{AS1}, \cite{AS11}, \cite{AS2}, \cite{ASDH} 
and for the discussions during the preparation of this text.  I thank J. Manschot, F. Nieri, C. Vafa and J. Walcher for many enlightening discussions that led to my understanding presented here in Section 4. I thank M. Gualtieri and Z. Komargodski for the discussions during the preparation of this manuscript and A. Gerasimov for the discussions of closely related topics over the years. My research was partly supported by the award of
the Simons Foundation to the Hamilton Mathematics Institute of the Trinity College Dublin
under the program ``Targeted Grants to Institutes''.

\vspace{2cm}

\section{Gauge fields and 2-cocycle}
\vspace{1cm}

Let $M_4={\mathbb{R}}^{1,3}$ be a four-dimensional space-time with local coordinates $(x_0,x_1,x_2,x_3)$ and Minkowski metric $ds^2=\sum_{\mu,\nu}\eta_{\mu\nu}dx^{\mu}dx^{\nu}, \eta^{\mu\nu}=diag(1,-1,-1,-1)$. Let $\bf{G}$ be a compact Lie group, $\mathcal{\bf{g}}$ its Lie algebra and $P \rightarrow M_4$ a principal $G$-bundle over $M_4$. We write $\mathcal{A}$ 
for the infinite-dimensional affine space of connections in the adjoint bundle $\text{ad} P=  P {\times_G} \mathcal{\bf{g}}$ 
and $\mathcal{G}_4=Map(M_4,G)$ for the infinite-dimensional group of gauge transformations. \footnote{When gauge field interacts with `matter' (for example fermions), one also uses the complex vector bundle $E \rightarrow M_4$ associated with a given finite-dimensional representation $\rho: G \rightarrow End V$ of $G$. We will use boldface notation ${\bf A}$ for the connections in the infinite-dimensional affine space of connections on $E$ (as well boldface notation for the parameters of gauge transformation) later in this text. } On the space of connections $\mathcal{A}$ gauge group acts as:

\begin{equation}
\label{act}
A \rightarrow A^g=gAg^{-1}+dgg^{-1}, \quad \quad A=\sum_{\mu=0}^3A_{\mu}dx^{\mu}  \in \mathcal{A}, \quad g \in \mathcal{G}_4,
\end{equation}
\begin{equation}
F \rightarrow F^g=gFg^{-1}.
\end{equation}
Curvature $F$ of a connection $A$ is defined as $F=(d-A)^2=-dA+A\wedge A$.
Corresponding Lie-algerbaic gauge transformations are:

\begin{equation}
\label{lieact}
A \rightarrow A^u=du-[A,u], \quad F \rightarrow -[F,u], \quad u \in \mathcal{\frak{g}}_4,
\end{equation}
where $\frak{g}_4$ denotes the Lie algebra of $\mathcal{G}_4$.

In the Hamiltonian approach for the four-dimensional gauge theories the Gauss law plays a central role. In this approach one considers space $M_3=\mathbb{R}^3$ 
(or any other three-dimensional space, but for simplicity we will treat here the case of ${\mathbb{R}}^3$), a time direction with coordinate $x_0=t$ and focus is on constant time slice (for example $t=0$). Now one has the principal $G$-bundle $P_3$ over $M_3$, connections (gauge fields) $A=\sum_{i=1}^3A_idx^i$ and time-independent gauge transformations $\mathcal{G}_3=Map(M_3,G)$. Infinite-dimensional phase space of this Hamiltonian system has coordinates $(A,E)$ where $E=\sum_iE_idx^i=\sum_iF_{0i}dx^i$. Denote by $t^a$ the generators of Lie algebra $\mathcal{\bf{g}}$ ($a=1,...,n=dimG; \text{Tr} (t^at^b)=-2\delta^{ab}, [t^a,t^b]=\sum_cf^{ab}_ct^c$) and write $A_i(x)=\sum_a A^a_i(x)t^a, E_i(x)=\sum_aE^a_i(x)t^a$. In these coordinates the canonical symplectic form on the phase space and canonical Poisson structure are given by: 

\begin{equation}
\label{canon}
\Omega=\int_{M_{3}} Tr \delta A_i \delta E_i, \quad \quad \{A^a_i(x), E^b_j(y)\}=\delta^{ab}\delta_{ij}\delta^{(3)}(x-y)
\end{equation}
and the Hamiltonian is ($B_i={1 \over 2} \sum_{jk}\epsilon_{ijk}F_{jk}$, $\epsilon_{ijk}$ - totally antisymmetric tensor, $\epsilon_{123}=1$):

\begin{equation}
\label{hamym}
H=\int_{M_3} Tr \sum_{i=1}^3 {{E_i^2+B_i^2} \over 2}
\end{equation}
%Yang-Mills field, the gauge field, can be associated with any compact semi-simple Lie group  $\bf{G}$. It is given by one-form $\sum_{\mu=1}^4A_{\mu}dx^{\mu}$ which takes value in the Lie algebra $\mathcal{G}$ of this group. 

Gauss Law generators are:

\begin{equation}
\label{gauss}
G(u)=\int_{M_3} Tr (\partial_iu-[A_i,u])E_i 
\end{equation}
They generate the Lie-algebraic gauge transformations \eqref{lieact} under Poisson brackets \eqref{canon}:

\begin{equation}
\label{poisgauge}
\{G(u), A\}=A^u=du-[A,u], \quad \quad \{G(u),E\}=E^u=-[E,u]
\end{equation}
and satisfy:

\begin{equation}
\label{infinlie}
\{G(u), G(v)\}=G([u,v])
\end{equation}
If one adds the matter, for example Weyl (or Dirac) fermions $\psi$ in the representation $\rho$, the phase space with its symplectic structure (Poisson brackets) and Hamiltonian will be modified appropriately, but one will always have the functional on phase space ${\bf G(u)}$ that generates gauge transformation and satisfies \eqref{infinlie} under the corresponding Poisson brackets in the Hamiltonian approach. In fact, in the case of fermions - the modified Gauss law ${\bf G(u)} = G(u)+\int_{M_3}Tr_{\rho}{\bf u} {\bf J}$ where ${\bf J}$ is a fermionic current, bilinear in $\psi$. 

Classical ``physical" phase space is given by a Hamiltonian reduction of above phase space under the contraints ${\bf G}({\bf u})=0$. Upon quantization, when ${\bf G(u)}$ becomes operator in an appropriate second-quantized Hilbert space and Poisson brackets are replaced by commutators, $\{\bullet,\bullet \} \rightarrow i[\bullet , \bullet]$ the relations \eqref{infinlie} might get modified, but ${\bf G(u)}$ is required to generate the gauge transformations \eqref{poisgauge} with Poisson bracket replaced by commutator and satisfy Jacoby identity. 

The study of non-abelian quantum anomalies in \cite{FS}, \cite{Mick} led to a possibility of the consistent modification of \eqref{infinlie} with a non-trivial extension given by a 
gauge-field dependent 2-cocycle:

\begin{equation}
\label{infinlienew}
[{\bf G(u)}, {\bf G(v)}]={\bf G([u,v])}+{i \over {12\pi^2}}\int _{M_3}Tr_{\rho} {\bf A}d{\bf u}d{\bf v}
\end{equation}
Here $Tr_{\rho}$, gauge field ${\bf A}$, and ${\bf u,v}$ all are in the representation $\rho$ of gauge group. 
%${\bf A} = \sum_a A^a \lambda^a, {\bf u}=\sum_a u^a \lambda^a, {\bf v}=\sum_a v^a\lambda^a$, $\lambda^a$ - generators of Lie algebra ${\bf g}$ in the representation $\rho$. 
This is the 2-cocycle Igor and I. Singer also found, but didn't publish, as mentioned in the introduction. 

It is important to stress that the extension in \eqref{infinlienew} is not
a central extension - the generators of the algebra, ${\bf G(u)}$, do not commute with it because $[{\bf G(u),A}]=d{\bf u}-{\bf [A,u]}.$ But direct check shows that the new commutation relations \eqref{infinlienew} satisfy Jacoby identity: 

\begin{equation}
\label{jac}
[{\bf G(u_1)},[{\bf G(u_2)},{\bf G(u_3)}]] + cyclic=0
\end{equation}
and that the extension is not trivial (in the space of local functionals of gauge field). 

If ${\bf G(u)}$ in quantum theory satisfies \eqref{infinlienew} instead of \eqref{infinlie} 
the theory is anomalous, the operator version of classical constraints ${\bf G}({\bf u})=0$ makes no sense - one needs to find a ``new way" to quantize the theory.
 
A few words about how this 2-cocycle appeared in the theory of quantum anomalies \cite{Adler}, \cite{BJ}. Here I will follow the paper \cite{FS}. One considers Yang-Mills field interacting with Weyl fermions $\psi$ in a representation $\rho$
of gauge group. A quantum anomaly is the assertion that the partition function $Z({\bf A})$ of Weyl fermions in an external gauge field $\bf A$ (which is formally given by 
 the determinant of Weyl operator - a ``square root'' of the determinant of the Dirac operator in the external gauge field $\bf A$) is not gauge invariant. One easily sees that there is a regularization (for example, using the $\zeta$-function) under which the determinant of the Dirac operator is in fact gauge invariant. Thus, the 
partition function of Weyl fermions $Z(\bf A)$ is invariant up to a phase:

\begin{equation}
\label{deter}
Z({\bf A}) \rightarrow  Z({\bf A}^g)=e^{-i\alpha_1(A,g)}Z(\bf A)
\end{equation}
One can modify the action of the group:

\begin{equation}
\label{grou}
Z(A) \rightarrow U(g)\cdot Z({\bf A})=e^{i\alpha_1(A,g)}Z({\bf A}^g)
\end{equation}
and under this new action $Z(\bf A)$ is gauge invariant: $U(g) \cdot Z(\bf A)=Z(\bf A)$ as it follows from \eqref{deter}.
Since \eqref{grou} needs to be a group action, $U(g_1)U(g_2) = U(g_1g_2)$, we conclude, when we act twice using
\eqref{grou}, that possible
$\alpha_1(A,g)$'s are restricted to those satisfying:

\begin{equation}
\label{onecoc}
(\delta\alpha_1)(A;g_1,g_2)=\alpha_1(A^{g_1},g_2)-\alpha_1(A,g_1g_2)+\alpha_1(A,g_1)=0 \quad \quad \quad mod \quad 2\pi
\end{equation}
%Lie algebra version of \eqref{deter} is:

%\begin{equation}
%\label{deterlie}
%G(\bf u)Z(\bf A)=\frak{a}(\bf A,u)Z(\bf A)
%\end{equation}

%\end{document}
Thus one is looking for the solution of \eqref{onecoc} in the space of (``local") functionals of gauge field $A$ and a group element $g$ up to the identification $\alpha_1(A,g) \rightarrow \alpha_1(A,g)+\delta\beta, \quad \delta\beta(A)=\beta(A^g)-\beta(A)$. \footnote{ Change of the regularization scheme corresponds to the redefinition of partition function $Z({\bf A}) \rightarrow e^{i\beta(A)}Z({\bf A})$ with some local functional $\beta(A)$ (local counterterm in physics), thus $\alpha_1(A,g)$ is defined up to the variation of this $\beta(A)$.} 

In general, one can introduce the "differential" $\delta$ in an abstract setup. Thus, let $G$ be an abstract group and $M$ a right $G$-module. Consider the complex $\alpha_{\bullet}(M,G)$ with cochains $\alpha_k(M,G)$, the spaces of functions on $M \times G^k$, k = 0,1,2,..., and with the differential $\delta$:

\begin{equation}
\label{coch}
\delta: \alpha_k(M,G) \rightarrow \alpha_{k+1}(M,G)
\end{equation}
defined by:

\begin{equation}
\label{defdelt}
(\delta\alpha_k)(m; g_1,g_2,...,g_{k+1})=\alpha_{k+1}(m;g_1,g_2,...,g_{k+1}) =
\end{equation}
$$=\alpha_k(m\cdot g_1;g_2,...,g_{k+1})-\alpha_k(m; g_1g_2,g_3...,g_{k+1})+$$
$$+\sum_{i=2}^n(-1)^i\alpha_k(m;g_1,...,g_i g_{i+1},g_{i+2},...g_{k+1})+
(-1)^{n+1}\alpha_k(m; g_1,...,g_k)
$$
One can check that $\delta^2=0$. This cochain was applied in \cite{FS} to the case when $G$ is the gauge group 
and $M=\mathcal{A}$ in order to interpret the anomalous phase $\alpha_1(A,g)$ (called Wess-Zumino-Witten action in physics) in \eqref{deter}  as a corresponding 1-cocycle since for $n=1$ \eqref{defdelt} gives exactly \eqref{onecoc}: 
$\delta\alpha_1(A,g)=0$. \cite{FS} suggested to use the bicomplex with operators $\delta$ and $d$
(the exterior differentiation operator arising in the theory of Chern-Simons secondary 
characteristic classes) in a following way: start with manifold $M^{2n+2}$ of dimensions $(2n+2)$ and consider the Chern polynomial 

\begin{equation}
\label{chern}
w_{2n+2}= {1 \over {n!}}tr_{\rho} F^{n+1}
\end{equation} 
Since
\begin{equation}
\label{desc}
dw_{2n+2}=0, \quad \quad \delta w_{2n+2}=0
\end{equation} 
one can find locally on $M^{2n+2}$ a differential form of degree $(2n+1)$ such that

\begin{equation}
\label{firstdesc}
d w_{2n+1}=w_{2n+2}
\end{equation}
According to Novikov, globally on $M^{2n+2}$ it is a multivalued form:

\begin{equation}
\label{Nov}
w_{2n+1}=d^{-1}w_{2n+2}
\end{equation}
but if $M^{2n+1}=\partial B^{2n+2}$ is a $2n+1$ dimensional cycle in $M^{2n+2}$ which is a boundary of a $(2n+2)$-dimensional submanfold $B^{2n+2}$ of $M^{2n+2}$ one can integrate $w_{2n+1}$ over $M^{2n+1}$ 
and obtain well-defined functional called Chern-Simons action:

\begin{equation}
\label{cs}
I_{CS}(A)=2\pi \int_{B^{2n+2}} w_{2n+2}(A')=2\pi \int_{M^{2n+1}} w_{2n+1}(A)
\end{equation}
where one chooses any extension $A'$ of the connection $A$ 
from $M^{2n+1}$ to $B^{2n+2}$. For two such extensions, the difference

\begin{equation}
\label{difference}
\int_{B^{2n+2}} w_{2n+2}(A')-\int_{B^{2n+2}} w_{2n+2}(A^{''})
\end{equation}
is the integral of $w_{2n+2}$ over a top-dimensional cycle in $M^{2n+2}$
so assuming the integrality of the cohomology class (the coefficient in the definition of
$w_{2n+2}$ was chosen appropriately)

\begin{equation}
\label{integrality}
[w_{2n+2} ]\in H^{2n+2}(M^{2n+2}, \mathbb{Z})
\end{equation}
one gets univalued exponent $e^{iI_{CS}(A)}$ and univaluedness of this exponent is 
enough (due to the path integral properties used in physics).

It is easily verifiable that the following $(2n + 1)$-form is closed:

\begin{equation}
\label{firststep}
\delta w_{2n+1} = w_{2n+1}(A^g)-w_{2n+1}(A); \quad \quad d\delta w_{2n+1}=d\delta d^{-1}w_{2n+2} =\delta w_{2n+2}=0
\end{equation}
It follows that, in contrast to the Chern-Simons action, its gauge variation

\begin{equation}
\label{varcs}
\alpha_1(A,g)= I_{CS}(A^g)-I_{CS}(A)
\end{equation}
is well defined (modulo $\mathbb{Z}$) also in the case when $M^{2n+1}$
 is not a cycle but a chain with boundary $\partial M^{2n+1} = M^{2n}$:
 \begin{equation}
 \label{firstone}
 \alpha_1(A,g)=2\pi \int_{M^{2n+1}} \delta w_{2n+1}(a)=2\pi \int_{M^{2n}}d^{-1} \delta w_{2n+1}=
 2\pi \int_{M^{2n}}d^{-1}\delta d^{-1} w_{2n+2}
 \end{equation}
and $\delta\alpha_1(A,g)=0$ as a consequence of $\delta^2=0$. 

Thus \eqref{firstone} gives the desired expression for the phase in \eqref{grou} (Wess-Zumino-Witten 
action) for the 
transformation property of the determinant of Weyl operator \eqref{deter} - in four-dimensional case of Weyl fermions 
we just choose $n=2$. 

Finally, if one considers gauge theory with $CS(A)$ action for the $2n+1$ dimensional space-time manifold with $2n$-dimensional boundary and with anomalous chiral Weyl fermions living on this boundary - the total partition function will be gauge invariant since variation of $CS$ action \eqref{varcs} will exactly cancel the anomalous phase of the chiral fermion partition function \eqref{grou}. This is frequently called the ``anomaly inflow" mechanism \cite{FS}, \cite{CH}.

The ``descent" procedure, described above, can be continued. At $k$-th step, when $d^{-1}$ is applied $k$-times, one gets a closed $(2n+2-k)$-form:

\begin{equation}
\label{kstep}
w_{2n+2-k}=\delta d^{-1}\delta d^{-1}...\delta d^{-1} w_{2n+2}
\end{equation}
which when integrated over appropriate boundary $\partial M^{2n+2-(k-1)} = M^{2n+2-k}$ produces $(k-1)$-cocycle: 

\begin{equation}
\label{kth}
\alpha_{k-1}(A;g_1,...,g_{k-1})=\int_{M^{2n+2-k)}}w_{2n+2-k},
\end{equation}

$$\delta\alpha_{k-1}=0$$

There is an algebraic version of this construction (equivalent to that of \cite{Zumino}, \cite{Stora} 
where Faddeev-Popov Ghosts were used and the BRST operator was defined instead of straightforward 
algebraic counterpart of $\delta$ from \eqref{defdelt}) - 
$k$-th cochain ${\bf a}_k(A; u_1,...,u_k)$ is a functional of gauge field and $u_1,...,u_k \in \frak{g}$ and $\delta$ in \eqref{defdelt} 
is defined
in terms of gauge transformation from Lie algebra - $A^u=du-[A,u]$, and group multiplication $g_1g_2$ is replaced by commutator $[u_1,u_2]$. In the algebraic version multi-valuedness, discussed above, is absent. 

The third step, $k=3$, leads to a 2-cocycle: 
\begin{equation}
\label{twococ}
\alpha_2(A; g_1,g_2)=\int_{M^{2n-1}} 
d^{-1}\delta d^{-1}\delta d^{-1}w_{2n+2}(A)
\end{equation} 
which may appear in the projective representation of gauge group:

\begin{equation}
\label{proj}
V(g_1) \cdot V(g_2)=e^{i\alpha_2(A;g_1,g_2)}V(g_1g_2)
\end{equation}
and the 2-cocycle property of $\alpha_2(A;g_1,g_2), \delta\alpha_2(A; g_1,g_2)=0$, guarantees associativity in \eqref{proj}.

Again, for $n=2$ one gets the 2-cocycle in three-dimensions, since the integral is over three-manifold in \eqref{twococ}. One can explicitly compute the algebraic form of this $\alpha_2(A;g_1,g_2)$ from the above construction (multivaluadness subtleties are absent in the algebraic setup, everything can be explicitly calculated starting descent from the Chern class $w_6={1 \over {3!}}tr_{\rho}F^3$) - the result is exactly the extension entering in the Gauss Law commutation relations given by the formula \eqref{infinlienew}.

In \cite{FS} it was conjectured that in the anomalous theory of Weyl fermions interacting with gauge fields in (3+1)-dimensions
the total space of quantum states is fibered over the space of all gauge fields $\mathcal{A}$ with fibers given by the fermionic Fock-space (denote the section by $\Psi(A)$) and gauge group acts as:

\begin{equation}
\label{totalspace}
V(g) \cdot \Psi(A)=U(A,g)\Psi(A^g)
\end{equation}
(it is useful to compare with \eqref{grou}) with some operator $U(A,g)$ satisfiying:

\begin{equation}
\label{gerb}
U(A,g)U(A^g,h)=e^{i\alpha_2(A;g,h)}U(A,gh)
\end{equation}
Unfortunately, for non-abelian $G$, such representation (in a proper Hilbert space)  was not found at that time. 
I have not been following the topic for over 30 years. I will be more than grateful if someone could
point me to a paper(s) in mathematics literature where (if) this has been done.
%but I believe mathematically rigorous construction of \eqref{infinlienew} is still unknown, though there have been some  developments. 
In the path integral approach the validity of
this conjecture was demonstrated in the 80s, see e.g. \cite{Madaj}.

\vspace{.5cm}
\section{Coadjoint orbits}
\vspace{1cm}

The first example of orbital integral of an exponential function goes back to Harish-Chandra 
(in modern days called Harish-Chandra-Itzykson-Zuber integral \cite{HCh}, \cite{IZ}):

\begin{equation}   \label{IZ}
\int_{U(n)} du \, e^{\rm Tr (A u B u^{-1})} =
C_n\frac{\sum_{\sigma \in S_n} (-1)^{|\sigma|} e^{\sum_i a_i b_{\sigma(i)}}}{\prod_{i<j} (a_i - a_j)(b_i-b_j)} 
\end{equation}
for two Hermitian $n \times n$ matrices $A$ and $B$ with  with distinct eigenvalues $a_1, \dots, a_n$  and $b_1, \dots b_n$. Here $du$ is the Haar measure on the group $U(n)$, $C_n=\prod_{i=1}^n i!$. The left hand side in the expression \eqref{IZ} can be understood, up to the inverse of Vandermond determinant $\Delta(B)={\prod_{i<j}(b_i-b_j)}$ of matrix $B$, as an integral over the coadjoint orbit $\mathcal{O}_B=\{ uBu^{-1}; u \in U(n)\}$ under the action of $U(n)$ by conjugations. Individual terms in the sum on the right hand side are in one-to-one correspondence with elements of the orbit invariant under conjugations by diagonal matrices given by permutations of eigenvalues of $B$: $B_\sigma=
{\rm diag}(b_{\sigma(1)}, \dots, b_{\sigma(n)})$ (this sum can be rewritten as a $\det \left [e^{a_ib_j} \right ]$).

Integral in \eqref{IZ} is an example of Duistermaat-Heckman (DH) integral \cite{DH}\footnote{One of the first papers expressing the Kirillov's integral over coadjoint orbits of compact Lie groups \cite{KirilB} as a sum \eqref{IZ}  was  \cite{STS}. }.
Let $M$ be a compact symplectic manifold of dimension ${\rm dim}_{\mathbb{R}}(M)=2d$ with symplectic form $\omega$. We assume that it carries a Hamiltonian action of a torus $T$ with isolated fixed points $p_1, \dots, p_m$. We denote the Lie algebra of $T$ by $\mathfrak{t}={\rm Lie}(T)$ (since fixed points are isolated weights of the $T$-action on  tangent spaces $T_{p_i}M$, denoted by $w_{a,i} \in \mathfrak{t}^*$ for $a=1, \dots, d, i=1, \dots, m$, are non-vanishing).  We denote the corresponding moment map by $\mu: M \to \mathfrak{t}^*$. The Duistermaat-Heckman formula computes the following oscillating integral
\begin{equation}
\label{osc}
I_M(\xi) = \int_M \frac{\omega^d}{d!} \, e^{i \langle \mu, \xi \rangle},
\end{equation}
where $\xi \in \mathfrak{t}$.
The result is given by the localization formula:
\begin{equation} \label{eq:DH}
I_M(\xi) =\left( \frac{2\pi}{\sqrt{-1}} \right)^d \sum_{i=1}^m \, \frac{e^{i \langle \mu(p_i), \xi \rangle} }{\prod_{a=1}^d \langle w_{a,i}, \xi\rangle } .
\end{equation}
which is known to be a scaling limit of localization formula for the $T$-equivariant index of a symplectic Dirac operator twisted by the pre-quantum line bundle.

For coadjoint orbits corresponding to compact Lie groups (orbit $\mathcal{O}_B$ in case of \eqref{IZ}) the $DH$ integral coincides with \eqref{IZ}. According to Kirillov, who originally introduced the orbit method in 1961 \cite{KirilB}, the character of the irreducible representation of the compact Lie group corresponding to the orbit ($\mathcal{O}_B$ in \eqref{IZ}) is given by \eqref{osc} (where $M$ is a coadjoint orbit) but up to a factor $p(\xi)$ which is universal and doesn't depend on orbit (and is identity at $\xi=0$) - see formula $\Phi$ in \cite{KirilB}. Kirillov also mentioned in \cite{KirilB} that proper integral formula for the character, which  automatically gives this pre-factor $p(\xi)$, should be the functional integral (Feynman integral, which explains why the main formula was denoted as $\Phi$) over coadjoint orbit and not an ordinary integral like in \eqref{osc} (thus he thought character should be obtained by quantum mechanics on coadjoint orbit), but such path integral was not known at that time; corresponding path integral for compact Lie groups $SU(N)$ and $SO(N)$ was obtained in 1980's, see \cite{AFS}.

Igor pioneered the topic of orbital integrals for the infinite-dimensional case. Basically he introduced the infinite-dimensional version of  \eqref{osc} for the coadjoint orbits of loop group $\widehat{LG}$ \cite{Fr1} suggesting the Wiener type measure (his integral probably best can be thought of in the spirit of the integration over the group manifold as in \eqref{IZ}). Calculating the integral he showed that answer can be written in a very similar form to \eqref{eq:DH}, see formula (53) below.  After Igor's work the orbit method for loop groups was developed in many papers, both in mathematics and physics literature. Here I list a few, in addition to those already mentioned. In physics paper \cite{Picken} was very similar to Igor's approach. Related, and very recently,  a new supersymmetric localization principle for the path integral over the tangent bundle to $LG$ was formulated in \cite{L1},  applied to derive an exact formula for the propagator on a group manifold as well as the Selberg trace formula \cite{L2}. Symplectic geometry approach to formal DH integral on $\widehat{LG}$ coadjoint orbits has been developed in \cite{We}, and more recently addressed by \cite{Jeffrey}.  The probabilistic approach to $\widehat{LG}$ orbital integrals has been further developed in \cite{De2}, \cite{De1}. 
 Rigorous mathematical derivation of DH integral for infinite-dimensional case of $\widehat{LG}$ coadjoint orbits is contained in the approach of \cite{B1}, \cite{B2} using hypoelliptic Laplacian on $G \times {\bf g}$. 
 
The orbit method appeared recently in a wide variety of topics in modern theoretical physics. To list just a few interesting connections, we mention the SYK model \cite{SY}, \cite{K} as studied in \cite{TV}-\cite{SW},  and the vacuum decay in CFT of \cite{PPT}. In the SYK model, and related generalizations, one is interested in $\widehat{\rm Diff}(S^1)$ coadjoint orbits. For example the infinite-dimensional orbital integrals of the type

\begin{equation}      
\label{SW}
I(\tau) = \int_{{\rm Diff}(S^1)} \mathcal{D}f \, \exp\left( \tau \int_{S^1} S(f(s)) ds\right)
\end{equation}
were recently studied in \cite{SW}, \cite{TW} (here $f(s)$ is a diffeomorphism $f(s) \in \widehat{\rm Diff}(S^1)$ and $S(f)$ is a 
Schwarzian derivative as a Hamiltonian on coadjoint orbit, see next paragraph) in relation to SYK.

As mentioned above - already Kirilliov in the 60s stressed that ``correct" approach to the relation between
characters and orbital integrals should be in terms of path integral. In quantum mechanics, for the phase space with local coordinates $(p,q) \equiv (p_i,,,p_n; q_i,...,q_n)$, Hamiltonian function $H(p,q)$ and Planck constant $\hbar$ 
the partition function is given by path integral (integral over closed paths, parametrized by $t$, of length $T$, on the phase space $(p(t),q(t))$ :

\begin{equation}
\label{pathi}
Z_{\hbar}=\int \prod_t\prod_i dp_idq_i e^{{i \over \hbar} \int (\sum_i p_i\dot{q}_i-H(p,q))dt}=\int \prod_t\prod_i dp_idq_i e^{{i \over \hbar} \int (d^{-1}\omega-H(p,q)dt)}
\end{equation}
In the limit $\hbar \rightarrow 0, T \rightarrow 0, T/\hbar=\tau=fixed$ (classical limit) this integral becomes an ordinary integral over phase space
as in \eqref{osc}:\footnote{This argument doesn't take into account certain normalization factors, but scaling property is captured properly, as shown by examples below.}

\begin{equation}
\label{pathii}
Z_{\hbar \rightarrow 0}=\int \prod_i dp_idq_i e^{-iH(p,q))\tau}
\end{equation}
thus assuming that character of irreducible representation, when phase space is given by coadjoint orbit, coincides with the partition function \eqref{pathi}, one expects \cite{AFS}, \cite{TV}, \cite{ASDH}, \cite{TW} that this classical limit will be given by corresponding $DH$ formula. 

This idea can be verified for the cases when
both character formula and $DH$ formula are known independently. It can be considered as a conjecture
when the answer for $DH$ integral is not known but character is known, e.g. from algebraic methods. One of the infinite-dimensional examples, when both the character and $DH$ integral formula are known, is precisely the case studied by Igor \cite{Fr1} which we will consider in a moment. 

Before 
considering the scaling for the infinite-dimensional case and Igor's formula we will illustrate this scaling property of the character formula on the examples of coadjoint orbits for the finite-dimensional compact Lie groups. Let $G$ be a compact Lie group. Consider a character $\chi_\lambda(g)$ of an irreducible representation of highest weight $\lambda$. There is a coadjoint orbit $\mathcal{O}_\lambda$ associated to this representation. We denote its dimension by ${\rm dim}(\mathcal{O}_\lambda)=2d$. Then, the character admits the following interesting asymptotic expansion:
\begin{equation}        
 \label{asymp_finite}
\chi_{k \lambda}\left(e^{\frac{h}{k}}\right) = \left( \frac{k}{2\pi} \right)^d \, I^{\rm DH}_{\mathcal{O}_\lambda}(h) + \dots
\end{equation}
Here $k \in \mathbb{Z}$ is a large parameter, $h \in \mathfrak{t}={\rm Lie}(T)$ is in the Lie algebra of the maximal torus $T \subset G$, $I^{\rm DH}_{\mathcal{O}_\lambda}(h)$ is a $DH$ integral formula \eqref{osc}, \eqref{eq:DH} for $M={\mathcal{O}_\lambda}$ ($h$ plays the role of $\xi$), and $\dots$ stands for subleading terms in $k$. Note that in \eqref{asymp_finite} we scale the highest weight $\lambda$ with a large factor $k$ and that at the same time we scale down the group element $\exp(h/k)$ so as it gets closer and closer to the group unit.

The first example we consider, in order to demonstrate the validity of \eqref{asymp_finite}, is the case of the character of the compact group $SU(2)$ and corresponding $DH$ integral (coadjoint orbit is a sphere $S^2_s$ of radius $s$ - radius corresponds to spin):

\begin{equation}
\label{su2}
\chi_s(\theta)={sin(s+{1 \over 2})\theta \over sin{\theta \over 2}}
\end{equation} 
and scaling \eqref{asymp_finite} gives:

\begin{equation}
\label{stwo}
I_{S^2}(\theta) = \lim_{k \to \infty} \frac{2 \pi}{k} \, \frac{ \sin\left(\left(ks+\frac{1}{2}\right)\frac{\theta}{k}\right) }{ \sin\left(\frac{\theta}{2k}\right) } =
2\pi \, \frac{ e^{is\theta} - e^{-is\theta} }{i\theta} = \int_{S^2_s} e^{i z \theta} dz\wedge d \phi 
\end{equation}
which is exactly the DH integral for the sphere $S^2_s$ expressed in cylindrical coordinates.

Similarly, for a coadjoint orbit $O_\lambda$ of a compact connected Lie group $G$ we obtain
the following scaling
\begin{equation}
\label{gener}
\lim_{k \to \infty} \left( \frac{2\pi}{k}\right)^d \, \sum_{w \in W} \, \frac{e^{i k\langle w(\lambda), \xi /k \rangle}}{\prod_{\alpha \in \Delta_+}
(1 - e^{-i \langle w(\alpha), \xi/k \rangle})} =
\left( \frac{2\pi}{\sqrt{-1}} \right)^d \sum_{w\in W} \frac{e^{i\langle w(\lambda), \xi \rangle}}{ \prod_{\alpha \in \Delta_+}\langle w(\alpha), \xi \rangle} .
\end{equation}
The left hand side is scaled character and the right hand side of this equation is the Duistermaat-Heckman localization formula for the orbital integral \eqref{eq:DH}. Thus the heuristic argument with path integrals in \eqref{pathi}, \eqref{pathii} turns out to be precisely correct with appropriate normalization factor as in \eqref{asymp_finite}.

Now we move on to the infinite-dimensional situation, the case of loop groups (we focus on the case of  $\widehat{{\rm su}(2)}_l$ for simplicity). For $\mathfrak{g}={\rm su}(2)$ and $l$ a positive integer, the Kac-Moody algebra $\widehat{{\rm su}(2)}_l$
admits $l+1$ integrable irreducible representations labeled by spins $j=0, 1/2, \dots, l/2$. The Kac-Weyl character formula gives characters of these representations:
\begin{equation}
\label{kw}
\chi_{l,j}(\tau, \gamma) ={\rm Tr}_{V_{l,j}}\left( e^{2\pi i \tau L_0+ i \gamma J_0^z}\right) = \frac{\Delta_{l,j}(\tau, \gamma)}{D(\tau, \gamma)},
\end{equation}
\begin{equation}
\label{num}
\Delta_{l,j}(\tau, \gamma)= e^{2\pi i \tau j(j+1)/(l+2)} \sum_{n \in \mathbb{Z}}
e^{2\pi i \tau((l+2)n^2 + n(2j+1))} ( e^{i(j+(l+2)n)\gamma} - e^{-i(j+1+(l+2)n)\gamma}),
\end{equation}
and the  denominator does not depend on $j$:
\begin{equation}
\label{dmntr}
D(\tau, \gamma) = (1 - e^{-i\gamma})\prod_{r=1}^\infty(1-e^{2\pi i r \tau})(1-e^{2\pi i r \tau} e^{i \gamma})(1 - e^{2 \pi i r \tau} e^{-i\gamma}).
\end{equation}

Inspired by the finite dimensional case, we introduce the following scaling:
\begin{equation} \label{kmscal}
l \mapsto kl, j\mapsto kj, \tau \mapsto \frac{\tau}{k}, \gamma \mapsto \frac{\gamma}{k}.
\end{equation}
Under scaling transformations,  the numerator of the character formula behaves as follows:
\begin{equation}
\label{dscal}
\Delta_{kl,kj}\left(\frac{\tau}{k}, \frac{\gamma}{k}\right) \rightarrow_{k\to \infty} e^{2\pi i \tau j^2/l} \sum_{n \in \mathbb{Z}}
e^{2\pi i \tau((ln^2 + 2j n)} ( e^{i(j+ln)\gamma} - e^{-i(j+ln)\gamma}).
\end{equation}
Note that it has a well defined limit, and that all the terms of the Kac-Weyl formula are still present with slightly simplified exponents.

Now we turn to the large $k$ behaviour of the denominator of the character formula.
Recall that the Dedekind $\eta$-function $\eta(\tau)= q^{1/24} \prod_{n=1}^\infty (1-q^n)$, where $q=\exp(2\pi i \tau)$ and ${\rm Im} \, \tau >0$ verifies the modular property
\begin{equation}
\label{ded}
\eta(-1/\tau) = \sqrt{-i\tau} \eta(\tau).
\end{equation}
This implies
\begin{equation}
\label{subs}
\frac{1}{\prod_{n=1}^\infty (1 - q^n)} = \frac{1}{\prod_{n=1}^\infty (1 - e^{2\pi i n \tau})} =
\frac{\sqrt{-i\tau} \, e^{\pi i(\tau + \tau^{-1})/12}}{\prod_{n=1}^\infty (1 - e^{-2\pi i n/\tau})} .
\end{equation}

Furthermore, recall the following identity:
\begin{equation}        \label{identity}
(1 - e^{-i\gamma})\prod_{l=1}^\infty(1-e^{2\pi i l \tau} e^{i \gamma})(1 - e^{2 \pi i l \tau} e^{-i\gamma})=(q^x,q)_\infty (q^{1-x}, q)_\infty,
\end{equation}
where  $(a,q)_\infty=\prod_{k=0}^\infty(1-aq^k)$ is the Pochhammer symbol and $x=-\gamma/2\pi \tau$.
The combination of Pochhammer symbols on the right hand side of \eqref{identity} has the following asymptotic behaviour  (see Corollary 3.3 in \cite{Lambert} and Corollary 1.3 in \cite{Katsurada}):
\begin{equation}
\label{pohone}
(q^x,q)_\infty(q^{1-x},q)_\infty \sim_{q \to 1} 2 \sin(\pi x) e^{\pi^2/3\ln(q)} q^{-1/2(1/6-x+x^2)} .
\end{equation}
Applying rescaling \eqref{kmscal}, we obtain the asymptotic behaviour of the inverse denominator for $k\to \infty$: 
\begin{equation}
\label{pohtwo}
D\left(\frac{\tau}{k}, \frac{\gamma}{k}\right)^{-1} \sim_{k\to \infty} \frac{1}{2 \sin(\gamma/2\tau)} \left( \frac{- i \tau}{k}\right)^{1/2} e^{i \pi k/4 \tau} =
\left( \frac{2\pi}{k} \right)^{1/2} e^{i\pi k/4 \tau} \cdot  \frac{(-i \tau)^{1/2}}{2 \sqrt{2\pi} \sin(\gamma/2\tau)}.
\end{equation}
The divergent factor is of the form $(k/2\pi)^d \exp(i \pi \alpha k/12 \tau)$ with $d=-1/2$ and $\alpha=3$. The factor  $(k/2\pi)^d$ is similar to the finite dimensional case even though the ``dimension'' $d$ is not an integer and is negative. The factor $\exp(i \pi \alpha k/12 \tau)$ has an essential singularity at $k=\infty$, and this singularity depends on the Lie algebra parameter $\tau$.\footnote{A possible CFT interpretation of the exponent $\alpha$ is the number of free fields in the theory. Indeed, irreducible representations of $\widehat{LSU(2)}$ can be resolved in terms of free field Wakimoto modules with 3 free fields. Thus, one might conjecture that $\alpha=dim G$ for higher rank groups.}

 In conclusion, the behavior of the character under the scaling transformation is given by tthe formula
\begin{equation}
\label{scaltwo}
\chi_{kl,kj}\left(\frac{\tau}{k}, \frac{\gamma}{k}\right) \sim_{k\to \infty} 
\left( \frac{2\pi}{k} \right)^{1/2} e^{i\pi k/4 \tau} 
 \,  \cdot  I_{l,j}(\tau, \gamma),
\end{equation}
where 
\begin{equation}
\label{finkm}
I_{l,j}(\tau, \gamma)= \frac{(-i \tau)^{1/2}}{2 \sqrt{2\pi} \sin(\gamma/2\tau)} e^{2\pi i \tau j^2/l} \sum_{n \in \mathbb{Z}}
e^{2\pi i \tau((ln^2 + 2j n)} ( e^{i(j+ln)\gamma} - e^{-i(j+ln)\gamma}).
\end{equation}
The expression \eqref{finkm}, appearing in the right hand side of scaling property \eqref{scaltwo}, is exactly the formula derived by Igor \cite{Fr1} as a loop group generalization of Kirillov's orbit integral (with some simple identification of parameters $\tau,\gamma$ with those used in \cite{Fr1}, see \cite{ASDH} for details). Exponential pre-factor, similar to one in \eqref{scaltwo}, also appears in the scaling relation between Virasoro character and $DH$ integral over $\widehat{\rm Diff}(S^1)$ coadjoint orbit \cite{SW} as shown in \cite{ASDH}.

%\end{ex}

\vspace{.5cm}
\section{$WZW_4$ and holomorphic $CS$}
\vspace{1cm}
$WZW_4$ theory is a four-dimensional avatar of the $WZW_2$ model.\footnote{During my lecture at Igor's 70th birthday conference due to the time limit I was not able to cover this part, I talked about it in \cite{Ssum} instead.} 
Let $X_4$ be a  four-manifold
equipped  with a
metric  $h_{\mu\nu}$ and a
closed 2-form $\omega \in \Omega^2(X;\mathbb{R})$.
Let $g\in Map(X_4, G)$ for a Lie group $G$.
Fix a reference field configuration $g_0(x)$.
Then, for any $g(x)$ in the same homotopy
class as $g_0(x)$ we may define
the Lagrangian:

\begin{equation}
\label{wzwiii}
S_\omega[g;g_0]=
{f_\pi^2\over  8 \pi} \int_{X_4}  {Tr} g^{-1} d g \wedge * (g^{-1} d
g)
+ {i \over  12 \pi} \int_{X_5} \omega\wedge {Tr}(g^{-1} d g)^3
\end{equation}

Here  $f_\pi$ is a dimensionful parameter. $Tr$ is in adjoint representation. 
In the integral over
 $X_5= X_4 \times I $ in \eqref{wzwiii}
$\omega$ is independent of the fifth
coordinate; moreover, we use
a homotopy of $g$ to  $g_0$. The action is independent of the
choice of homotopy up to a multiple of the periods of $\omega$.
%\foot{Note that $X_5$ is a cylinder, rather than a cone. This is necessary since $X_4$ might not be cobordant to zero, and since the periods of $\omega$ might be nontrivial.  As was noted in \etingof, the latter fact caused difficulties in finding a ``Mickelsson''-type construction \mick\  of $\widehat {\rm Map}({\Sigma}, G)$ where $\Sigma$ is a Riemann surface. Using a cylinder and a homotopy construction this problem can be overcome  \cenexts.  A completely different solution to this problem  has recently been described in \fk.}
The classical equations of motion, following from \eqref{wzwiii}, are:
\begin{equation}
\label{gcem}
 d * g^{-1} d g + \omega \wedge g^{-1} dg \wedge g^{-1} dg = 0
 \end{equation}
 %\begin{equation}
% \label{gcemo}
 %\partial_{\mu}
%h^{\mu\nu} \sqrt{h}
%g^{-1} \partial_{\nu} g  + \epsilon^{\alpha \beta \gamma \delta}
%\omega_{\alpha \beta} g^{-1} \partial_{\gamma} g g^{-1} \partial_{\delta} g
%=0
%\end{equation}
When 
$X_4$ is a
complex four-manifold with
K\"ahler metric $
\omega = {i  \over  2} f_\pi^2 h_{i \bar j} dz^i \wedge d z^{\bar j}
$
one gets the point in the space of Lagrangians, the ``K\"ahler point.''
Now one can rewrite the action as:
\begin{equation}
\label{wzwiiv}
S_\omega[g]=
-{i \over  4 \pi} \int_{X_4}
\omega\wedge  {Tr} \bigl(g^{-1} \partial g \wedge g^{-1} {\bar \partial} g\bigr)
+ {i \over  12 \pi} \int_{X_5} \omega\wedge {Tr}(g^{-1} d g)^3
\end{equation}

The equations of motion following from
\eqref{wzwiiv} are:
\begin{equation}
\label{wzwv}
\omega\wedge{\bar \partial} (g^{-1}\partial g) = 0 
\end{equation}
which means that (2,1)-form current $J^{(2,1)}=\omega \wedge g^{-1} \partial g$ is $\bar \partial$ closed. At this special K{\"a}hler point $WZW_4$ theory is equivalent to the classical field theory of the N = 2 open strings for any group $G$ \cite{OV}, \cite{Ber}, \cite{LMNSavatar}, \cite{LMNStwo}. \footnote{String investigations of this theory have focused on the S-matrix for $\pi$, defined as $g=e^{i\pi}$,  hence one may expect the connections with the nonperturbative N = 2 string theory.}

From now on we will only focus on K\"ahler point. The coefficient of the WZ term is quantized.
Two different homotopies to
$g_0$  define a map
$ g: S^1\times X_4 \to G$.
Thus, if the group $G$ is
non-abelian,
 the measure $\exp i S$ in the
path integral is only
well-defined if
\begin{equation}
\label{quanti}
\omega\wedge
{1 \over  12 \pi} {Tr}(g^{-1} d g)^3\in H^5(S^1\times X_4;2\pi \mathbb{Z})
\end{equation}
which forces the cohomology class
 $[\omega]$ to lie in the lattice:
\begin{equation}
\label{contiii}
[\omega] \in H^2(X_4; \mathbb{Z})
\end{equation}
The class $[\omega]$ is the
four-dimensional analog of level $k$ in $WZW_2$.
Note that although $[\omega]$ is quantized
the Lagrangian depends on the representative
of the class.
Since $\omega$ is of type $(1,1)$
condition \eqref{contiii}  implies
$[ {\omega}] \in H^2(X;\mathbb{Z})\cap H^{1,1}(X;\mathbb{R})$
so the metric is Hodge, and,
by the Kodaira embedding theorem,
if $X_4$ is compact, it
must be  algebraic.

The important property of $WZW_4$ theory is that at the K\"ahler point quantum theory is finite \cite{LMNSavatar} (this is a very special point - quantum theory can not be deformed, away this point it  becomes non renormalizable).  The argument is very similar to that of in $WZW_2$ - first one demonstrates it is finite
in one-loop by calculating one loop determinant in the background field method, after which one uses the holomorphicity property of
$(2,1)$ current $J^{(2,1)}$ in \eqref{wzwv} to claim it must be finite to all loop orders (like in $WZW_2$ this holomorphicity will be violated if there are divergences at higher loops). This all loop finiteness statement is also well supported by studies of perturbative amplitudes in open $N=2$ strings (self-dual YM) which show no loop divergences.

Quantum $WZW_4$ theory has infinite-dimensional global, two-loop, symmetry $g(z,\bar z) \rightarrow h_L(\bar z)g(z,\bar z) h_R(z)$ (not to be confused with gauge symmetry) and this allows to calculate many correlation functions (of surface operators); 1-cocycle properties (from Section 2) of the $WZW_4$ action play a key role in the quantum theory. Conservation law, holomorphicity of $(2,1)$ current $J^{(2,1)}$ in \eqref{wzwv}, again leads to useful Ward identities. The five-dimensional viewpoint (``K\"ahler-Chern-Simons" theory), similar to 3d $CS$/$WZW_2$ relation, allows to define and calculate ``four-dimensional Verlinde formula" as well as establish the connection to Donaldson theory, etc. All these can be found in \cite{LMNSavatar}. 

One needs to emphasize again that $WZW_4$ theory is not a gauge theory. Rather it is a very special sigma-model, and its infinite-dimensional symmetries are not gauge symmetries - they are global symmetries of the theory (similar to the infinite-dimensional symmetries of $WZW_2$ which are global symmetries there too). Coupling of $WZW_4$ to gravity is also a well-defined quantum theory for any group and is equivalent to open/closed $N=2$ strings \cite{OV}, \cite{LMNStwo} -  $WZW_4$ is a well-defined theory, with or without coupling to the dynamical gravity and for any group $G$.

A bit of history - the Lagrangian \eqref{wzwiiv} was first written in \cite{Don}, 
the equations of motion \eqref{wzwv} appeared even before in \cite{Yang} as being equivalent to self-dual Yang-Milles equations (for Hermitian $g$ and after solving part of the self-dual YM equations) and are often called Yang equations. It was studied in \cite{Nair1}, \cite{Nair2} as a natural generalization of the $2d$ CFT/ 3d $CS$ correspondence. 

The equations of motion \eqref{wzwv} are integrable and can be solved using the twistor transform (see brief description in \cite{LMNSavatar}, page 3). Actually, as explained in \cite{Ward}, \cite{AW}, instantons on $\mathbb{S}^4$ correspond to holomorphic bundles  $F^{0,2}(A)={\bar \partial} \bar A + {\bar A}^2=0$ on $\mathbb{CP}^3$ (non-compact version relates equations on $\mathbb{R}^4$ and twistor space $\mathbb{PT}$): 

\begin{equation}
\label{AWard}
{\hat x}=q_2^{-1}q_1, \quad {\hat x}=x^0+i\sum_{i=1}^3x^i\sigma^i, \quad q_1 = \left( \begin{array}{ll}
z_1 & z_2 \\
- {\bar z}_2 & {\bar z}_1
\end{array}
\right), \quad q_2=\left(\begin{array}{ll}
z_3 & z_4 \\
-{\bar z}_4 & {\bar z}_3
\end{array}
\right)
\end{equation}
where $\sigma^i$ are Pauli matrices, $x \in \mathbb{R}^4$, $z \in \mathbb{C}^4$ and $q_i$ are quaternions. Now one can check that if gauge field $a=\sum_{\mu}a_{\mu}dx^{\mu}$ is instanton, satisfies self-duality condition $F(a)=\star F(a)$, then $A_i(z, {\bar z})=\sum_{\mu}a_{\mu}(x){dx^{\mu} \over dz^i}$ satisfies $F^{0,2}(A)=0$ (in fact - this is an equation in $\mathbb{CP}^3$ and not $\mathbb{C}^4$ due to the scale invariance of \eqref{AWard} under $q_i \rightarrow \lambda q_i$). The reverse, exactly how to obtain the instanton on $\mathbb{S}^4$ using this equivalence, is done through $ADHM$ construction \cite{ADHM}.

Let us stress two moments here: a) solutions to self-dual YM equations on $\mathbb{S}^4$ modulo its gauge transformation have been mapped to the solutions of $F^{0,2}(A)=0$ on $\mathbb{CP}^3$ modulo its gauge transformations, b) Yang equation \eqref{wzwv} is a result of solving the parts of the self-duality equations ($F^{2,0}(a)=0$ and $F^{0,2}(a)=0$ parts) and consecutively the final equation \eqref{wzwv} has no remaining gauge symmetry. 

Now let me turn to more recent developments. The relation between equation \eqref{wzwv} on $\mathbb{R}^4$ and $F^{0,2}(A)=0$ on twistor space $\mathbb{PT}$, which is true for any group $G$, can be written as a classical equivalence between action \eqref{wzwiiv} and holomorphic $CS$ action on $\mathbb{PT}$:

\begin{equation}
\label{hcs}
S_{HCS}= \int_{X} \Omega \wedge Tr({\bar A} \wedge {\bar \partial} {\bar A}+ {2 \over 3} {\bar A} \wedge {\bar A} \wedge {\bar A})
\end{equation}
$X=\mathbb{PT}$ and $Tr$ is again in the adjoint representation. One can consult \cite{Costello},  \cite{Costelloo}, \cite{Costellotwo}, \cite{Skinner} for detailed definition of \eqref{hcs} on $\mathbb{PT}$ (including that of holomorphic $(3,0)$-form $\Omega$ on $\mathbb{PT}$ which is used here - actually meromorphic form is used, poles are allowed and their boundary contributions are eliminated by imposing appropriate conditions on the gauge field so equations of motion derived from \eqref{hcs} are $F^{0,2}=0$, and the action \eqref{hcs} is well-defined as integral converges over $X=\mathbb{CP}^3$). In these papers one can also find the explicit formulae relating $g$ in \eqref{wzwiiv} and $\bar A$ in \eqref{hcs} such that the actions \eqref{hcs} and \eqref{wzwiiv}  become identical after solving the part of the six-dimensional equations for unphysical modes. 

Here we need one important element of the above classical ``off-shell" equivalence between $WZW_4$ in 4d and holomorphic $CS$ in 6d. In establishing this equivalence the twistor space $\mathbb{PT}$ is viewed as a $\mathbb{CP}^1$ bundle over $\mathbb{R}^4$ and the six-dimensional theory is treated as a four-dimensional theory on $\mathbb{R}^4$ with infinitely many fields with only zero-modes being physical - other modes turn out to be unphysical, either auxiliary fields not  containing  space-time derivatives along $\mathbb{R}^4$ or ``non-propagating" degrees which are guage transformations (see \cite{Costellotwo}). All these unphysical degrees of freedom are classically ``integrated out" (by solving the equations of motion for these fields and substitutes back into the action)  or eliminated by the gauge transformations, after which the actions \eqref{hcs} and \eqref{wzwiiv}  become identical. 

Let's stress once more - in this ``off-shell" equivalence it is very important that some fields (auxiliary/unphysical from 4d point of view) are classically consistently eliminated and in that the gauge invariance of classical action (equations of motion) is used.

Thus, if we collect the information presented so far, the conclusion is that there are several classical equivalences (with non-dynamical gravity, in the fixed gravitational background) for any group $G$, on the level of classical actions together with the equations of motion, and two of them involve $WZW_4$ (the equivalence which doesn't involve $WZW_4$, $2^{nd}$ one below, is important here since it is the one which leads to a puzzle we would like to discuss). These equivalences are the main reason of current presentation:

{\bf I.} between $WZW_4$ and $N=2$ open strings (self-dual Yang-Mills as mentioned before), 

{\bf II.} between holomorphic $CS$ theory on (compact) Calabi-Yau manifold $X$ in \eqref{hcs} and open topological strings \cite{WCS}, 

{\bf III.} between holomorphic $CS$ theory on twistor space $\mathbb{PT}$ and $WZW_4$ on $\mathbb{R}^4$ (or in compactified version - between holomorphic $CS$ on $\mathbb{CP}^3$ and $WZW_4$ on $\mathbb{S}^4$).

All three of these equivalences have a version with dynamical gravity since all of them have the interacting open/closed string theory counterparts. 

At the quantum level the situation with these equivalences is more complicated. I was discussing the corresponding issues, quantum equivalences both without and with dynamical gravity, in the summer 2022, around the time of Igor's '70th conference, with F. Nieri, J. Manschot, C. Vafa and J. Walcher \cite{MNVW}.
The conclusion is that:  {\bf Equiv I} holds on quantum level (both without and with dynamical gravity), this is an old and established fact \cite{OV}, {\bf Equiv II} holds on quantum level only when coupled to the dynamical gravity and for a very special gauge groups \cite{Walcher}, \cite{Oo}, \cite{Costello}, \cite{Costelloo}, \cite{Costellotwo}, with  {\bf Equiv III} there are some serious difficulties on quantum level. 
Below I comment on {\bf Equiv II} and {\bf Equiv III} based on these conversations and my current understanding. 

{\bf Equivalence II.} In topological string theory target space is Calabi-Yau manifold, so $X=CY$, thus integration over $X=\mathbb{PT}$ ($\mathbb{CP}^3$) in \eqref{hcs} needs to be replaced by the integration over $CY$ and $\Omega$ now is CY holomorphic 3-form in some fixed complex structure. In order to properly couple it with background gravitational fields one can deform this complex structure replacing ${\bar \partial} \rightarrow {\bar \nabla}={\bar \partial} - \mu^i \partial_i$ where $\mu^i$ is Beltrami differential: $\mu^i\partial_i=\mu^i_{\bar j} dz^{\bar j}\partial_i$.
After this \eqref{hcs} will depend on two classical closed string backgrounds $\Omega$ and $\mu$ which must satisfy
(on-shell) equations of motion of closed topological strings: 
%\begin{equation}
%\label{closedtop}
$ {\bar \nabla} \Omega+\partial_i\mu^i\Omega=0$ and ${\bar \partial}\mu^{i}-\mu^j\partial_j\mu^i=0$,  
%\end{equation}
meaning that $\Omega$ is holomorphic in the complex structure associated with $\mu^i$, and this $\mu^i$ defines the integrable complex structure satisfying the Kodaira-Spencer equation. One can write the action \eqref{hcs} on CY manifold in an arbitrary closed string background (off-shell) by introducing many auxiliary fields \cite{Imb}. And, these background fields can be made dynamical - in this case one gets the open/closed topological strings ($B$-model), holomorphic $CS$ coupled to Kodaira-Spencer gravity. 

It was pointed out already in \cite{Walcher}, from the world sheet considerations, that open topological string at the quantum level has anomalous behavior that needs to be taken care of.  This is an internal anomaly and such anomalous theories are usually not consistent quantum-mechanically. Solution proposed in \cite{Walcher} involves coupling to closed strings (gravity) after the orientifold projection and cancelling the orientifold charge by D-branes.\footnote{As explained by \cite{SinVafa} , and used in \cite{Walcher}, orientifold plane carries a D-brane charge which needs to be cancelled by D-branes to get the anomaly free theory. The focus of \cite{Walcher} was the A-model, but when translated to the B-model by mirror symmetry/T-duality, the \cite{SinVafa} charge that needs to be cancelled is $2^{\rm  d/2}$ ($d$ is the real critical dimension of the target space) and this implies that non-anomalous gauge group for 6-dimensional $CY$ must be $SO(8)$, although this was not written explicitly in \cite{Walcher}.} 

This topic was long forgotten until the space-time theory was studied in \cite{Costello}, \cite{Costelloo}, \cite{Costellotwo} where the proper and careful treatment of the space-time theory was given. Because of the chiral origin of the open string sector of the space-time theory (\eqref{hcs}, which depends only on $\bar A$), the proper definition of quantum theory suffers anomalies, similar to those described in Section 2 for chiral fermions. Again, here for the properly defined path integral over the open string modes ($\bar A$ and all other auxiliary fields mentioned above) in a fixed gravitational background ($\Omega, \mu$ and corresponding gravitational auxiliary fields) at the one loop one finds the anomaly \cite{Costello}, \cite{Costelloo}, \cite{Costellotwo}. There is an important difference between the chiral fermions in the background gauge field (where anomaly is ``harmless", unless these background gauge fields are replaced by the dynamical gauge fields) and the holomorphic $CS$ in a fixed gravitational background. Latter is a non-linear self-interacting theory and such anomaly would be again (as in the world-sheet approach mentioned above) an ``internal anomaly" leading to the inconsistent quantum theory at the higher loops. The explicit formula for this chiral anomaly, as always, can be obtained by the same cohomological descent procedure as in the Section 2 giving now the 1-cocycle with the values in the space of functionals of $\bar A$, $\alpha_1({\bar A}, u)$ (${\bf a}({\bar A},u)$ for algebraic 1-cocycle/anomaly). Since now we are in six-dimensions we need to start the descent from $w_8=Tr F^4$ ($Tr$ in the adjoint representation) and look for the holomorphic representatives, $\alpha_1({\bar A},g)$ holomorphic in gauge field (in some fixed complex structure, or holomorphic both in $\bar A$ and $\mu$ if one varies the complex structure). Actually, in order to see the anomaly, it is enough to consider the algebraic descent which leads (for a fixed complex structure) to ${\bf a}_1({\bar A},u)=
\int_X Tr(u(\partial {\bar A})^3)$.  This doesn't vanish. So, holomorphic $CS$ theory in non-dynamical gravitation background is anomalous and inconsistent.  

As in the world-sheet approach to topological strings, described above, the space-time theory can be made consistent quantum theory if coupled to the dynamical gravity (closed strings). Coupling to the dynamical gravity opens the opportunity to get the anomaly free theory if one can show that one loop quantum anomaly from the holomorphic $CS$ can be cancelled by the classical transformations from the gravity sector.  Such mechanism was found in \cite{Costello}, \cite{Costelloo}, \cite{Costellotwo} where it was shown explicitly that if the gravitational fields are made dynamical (thus properly coupling the holomorphic $CS$ to the appropriate Kodaira-Spencer theory) this is indeed the case - anomaly is cancelled from the contributions arising due to the transformations of gravitational fields via Green-Schwarz type mechanism, but only for the choice of the gauge group $G=SO(8)$ in \eqref{hcs}. 

Thus, following \cite{Costello}, \cite{Costelloo}, \cite{Costellotwo}, we conclude that the only consistent, anomaly free, six-dimensional quantum theory is the holomorphic $CS$ theory coupled to the dynamical gravity (open/closed topological string theory) with the gauge group $G=SO(8)$. Purely holomorphic $CS$ with non-dynamical gravitational background is inconsistent on quantum level. Simply stated -  {\bf Equiv II} can be extended to the quantum level only when a) properly coupled to the dynamical gravity and b) only for the gauge group $SO(8)$.

One may ask - does this conclusion lead to the consequences for the {\bf  Equiv III}? Does this mean that quantum 
$WZW_4$ is also inconsistent unless it is coupled to the dynamical gravity (coupled open/closed $N=2$ string theory)
and only for the gauge group $SO(8)$? If so - why such restriction was not found in 90s when studying $N=2$ strings, $WZW_4$, etc?

{\bf Equivalence III.} In order to address this relation one needs to replace $X=CY$, discussed in {\bf Equiv II}, by non-Calabi-Yau $X=\mathbb{PT}$ (or $\mathbb{CP}^3$). In the six-dimensional quantum 
theory, defined by the action functional \eqref{hcs}, in the non-dynamical gravitational background such replacement will not affect the anomaly, so quantum theory again will be consistent only for $G=SO(8)$ and only after coupling to the dynamical gravity. 

Thus, assuming {\bf  Equiv III} is true on quantum level, one may conclude that the same restrictions should apply to quantum $WZW_4$: this 4d theory is consistent quantum field theory only when properly coupled to the dynamical gravity and only for $G=SO(8)$. 

This conclusion creates a puzzle. Classical theory on $\mathbb{PT}$ ($\mathbb{CP}^3$) was equivalent to four-dimensional $WZW_4$ theory on $\mathbb{R}^4$ ($\mathbb{S}^4$) and this four-dimensional theory, as we stressed above, doesn't require any kind of restrictions from the previous paragraph on quantum level.  

What is the resolution of the puzzle? Most likely resolution is that at the quantum level the {\bf Equiv III} fails. Correct logic should be - instead of claiming that the restrictions from {\bf Equiv II} imply that 4d theory should be also restricted (by the coupling to gravity and $G=SO(8)$), the logical conclusion should be that there is no quantum {\bf Equiv III} even if we couple 4d $WZW_4$ theory to dynamical gravity (except for the coupled theory and group $G=SO(8)$ when one might certainly expect such equivalence to exist). 

This conclusion is hidden in the details of exactly how the classical equivalence was established. Mainly, from thinking about the classical theory on $\mathbb{PT}$ as a four-dimensional theory on $\mathbb{R}^4$ with the infinitely many unphysical degrees of freedom associated to $\mathbb{CP}^1$ (which were integrated out and consistently eliminated, classically).  For generic group $G$ we know that six-dimensional theory is anomalous - not gauge-invariant already at one loop. This in particular means that in the six-dimensional quantum theory above unphysical degrees of freedom (``longitudinal", gauge transformation, degrees of freedom) quantum-mechanically are not unphysical anymore - they are propagating degrees of freedom in quantum theory (as is always the case in theories with the internal quantum anomalies) and can not be consistently integrated out/eliminated in the same simple way as it was done classically. So, we can not establish quantum equivalence anymore in the same fashion as was done classically - when describing the classical {\bf Equiv III} we stressed the importance of the gauge symmetry for such equivalence to exist.

To reiterate - for the classical {\bf Equiv III} it was crucial that we solved the part of the classical equations of motions in six-dimensions eliminating auxiliary and unphysical degrees of freedom; thus, since six-dimensional theory is anomaly free only when coupled to dynamical gravity and only for $SO(8)$ gauge group,  integrating out the same degrees of freedom (as in classical theory) in quantum theory is a harmless operation only for the holomorphic $CS$ theory coupled to dynamical gravity and only for the $SO(8)$ gauge group - {\bf Equiv III} holds on quantum level only for the coupled theory and $G=SO(8)$. For other groups, coupled to dynamical gravity or not, one can not eliminate consistently  these degrees of freedom in quantum theory, so for other groups {\bf Equiv III} fails. Important conclusion - $WZW_4$ does not suffer any problems if $G$ is not $SO(8)$. At least these problems can not arise from those in $6d$ for such groups, since 4d theory is no longer equivalent to 6d theory, the restrictions in 6d theory have no consequences for the existence (or non-existence) of quantum theory in 4d.

For Igor and I, as I already mentioned in the introduction, the study of $WZW_4$ theory was a spinoff of our attempts to find the quantum field theory interpretation of Nakajima's results.
We have had many other spinoffs that led to many interesting developments - I only covered $WZW_4$ here since this topic has recently seen new important developments.

\end{document}